\documentclass{article}

\PassOptionsToPackage{numbers, compress, longnamesfirst, comma}{natbib}

\usepackage[final]{neurips_2022}

\usepackage[utf8]{inputenc} 
\usepackage[T1]{fontenc}    
\usepackage{hyperref}       
\usepackage{url}            
\usepackage{booktabs}       
\usepackage{amsfonts}       
\usepackage{nicefrac}       
\usepackage{microtype}      
\usepackage{xcolor}         
\usepackage{graphicx}
\usepackage{multirow}
\usepackage{multicol}
\usepackage{longtable}

\title{FMG-Net and W-Net: Multigrid Inspired Deep Learning Architectures For Medical Imaging Segmentation}

\author{%
  Adrian Celaya \\
  Rice University\\
  Houston, TX 77005 \\
  \And
  Beatrice Riviere \\
  Rice University\\
  Houston, TX 77005 \\
  \And
  David Fuentes \\
  The University of Texas MD Anderson Cancer Center\\
  Houston, TX 77003\\
}

\begin{document}

\maketitle

\begin{abstract}
Accurate medical imaging segmentation is critical for precise and effective medical interventions. However, despite the success of convolutional neural networks (CNNs) in medical image segmentation, they still face challenges in handling fine-scale features and variations in image scales. These challenges are particularly evident in complex and challenging segmentation tasks, such as the BraTS multi-label brain tumor segmentation challenge. In this task, accurately segmenting the various tumor sub-components, which vary significantly in size and shape, remains a significant challenge, with even state-of-the-art methods producing substantial errors. Therefore, we propose two architectures, FMG-Net and W-Net, that incorporate the principles of geometric multigrid methods for solving linear systems of equations into CNNs to address these challenges. Our experiments on the BraTS 2020 dataset demonstrate that both FMG-Net and W-Net outperform the widely used U-Net architecture regarding tumor subcomponent segmentation accuracy and training efficiency. These findings highlight the potential of incorporating the principles of multigrid methods into CNNs to improve the accuracy and efficiency of medical imaging segmentation.
\end{abstract}

\section{Introduction}
Medical imaging segmentation is an essential task in medical imaging analysis that involves dividing an image into regions of interest based on anatomical or pathological features. It is crucial in various clinical applications, including disease diagnosis, treatment planning, and surgical guidance. Accurate segmentation is critical for precise and effective medical interventions and can significantly impact patient outcomes \cite{intro-2, intro-3}.

Deep learning methods, specifically convolutional neural networks (CNNs), have emerged as powerful tools for medical imaging segmentation \cite{unet, unet-3d, nnunet, unet-common}. CNNs have remarkably succeeded in accurately segmenting anatomical structures and pathological regions from medical images. However, despite these advancements, CNNs still face challenges in preserving fine structures and handling significant variations in image scales \cite{antonelli2022medical, kitrungrotsakul2017robust, lim2018foreground}. These challenges are particularly evident in complex and challenging segmentation tasks, such as the BraTS multi-label brain tumor segmentation challenge. In this task, accurately segmenting the various tumor sub-components, which vary significantly in size and shape, remains a significant challenge, with even state-of-the-art methods producing substantial errors \cite{brats-nnunet, raza2023dresu, cao2023mbanet}.

Geometric multigrid methods (GMMs) are well-established techniques used in scientific computing to efficiently solve problems with multiple scales \cite{zhu2006multigrid, riviere2008discontinuous, shah1989analysis}. These methods use a hierarchy of grids with decreasing resolutions to efficiently capture large and fine-scale features. By capitalizing on these principles, we proposed two novel deep learning architectures, FMG-Net and W-Net, which incorporate the well-established ideas of the full and W-cycle multigrid methods into CNNs for medical imaging segmentation. These architectures aim to address the limitations of CNNs in handling fine-scale features and variations in image scales by incorporating the principles of these GMMs into the network architecture. Additionally, we aim to take advantage of the convergence properties of the full and W-cycle networks to reduce the number of iterations (i.e., epochs) required to achieve satisfactory results. The proposed architectures show promising results in accurately segmenting brain tumors while converging to lower loss function values in fewer epochs than the popular U-Net architecture, demonstrating their potential to improve medical imaging segmentation and ultimately enhance patient care.

\subsection{Background and Previous Work}
GMMs are a class of iterative methods that arise in scientific and engineering applications for solving linear systems of equations of the form $Au = f$, where the linear system $A$ and the unknown variable $u$ relate to a geometric grid involving multiple resolutions \cite{multigrid, saad2003iterative}. Intrinsic to these methods is the construction of a series of grids of appropriate resolution. At each resolution, the constructed grid allows for the specific frequencies to be resolved, without aliasing. Like GMMs, many common CNN architectures for imaging tasks rely on manipulating images (or image features) at multiple scales because natural images encapsulate data at multiple resolutions. As a result, most CNN architectures -- including many popular state-of-the-art methods such as nnUNet \cite{nnunet} --  follow a pattern of downsampling and upsampling, following the intuition of the original U-Net paper \cite{unet} that popularized this approach. This common idea of hierarchical grids seen in GMMs and CNNs has guided the design of recent CNN architectures, resulting in smaller networks (in terms of the number of parameters) that achieve similar performance to current methods. A brief survey of these methods is provided below. 

In their 2019 paper \cite{mgnet}, He and Xu initially explored the similarities between multigrid methods and CNNs. Their approach was to leverage the principles of multigrid methods to create a unified framework, which they called MgNet. This framework simultaneously recovered some CNNs for image classification and GMMs, leading to a better understanding of the functions of various convolution operations and pooling used in CNNs. This analysis resulted in the development of modified CNNs with fewer weights and hyperparameters, which exhibited competitive and sometimes better performance when applied to CIFAR-10 and CIFAR-100 datasets. However, the applications of this approach were limited to image classification models, and no extension was made to segmentation.

Celaya et al. utilize the nearly identical structures of the V-cycle multigrid method and the U-Net segmentation architecture to propose the PocketNet paradigm for deep learning models, a straightforward modification to existing architectures that substantially reduces the number of parameters and maintains the same performance as the original architecture \cite{pocketnet}. This modification questions the long-held assumption that doubling the number of features after each downsampling operation (i.e., pooling or convolution) is necessary for convolutional neural networks \cite{lenet, imagenet, alexnet, unet-3d, unet}. However, their results apply the concepts of GMMs to modify existing architectures similar to the V-cycle multigrid method (i.e., U-Net and nnUNet). Two other well-known GMMs are the full multigrid (FMG) and W-cycles, which construct a more complex hierarchy of grids than the V-cycle and are not utilized in their work.

\subsection{Novel Contributions}
In this paper, we make the following novel contributions:
\begin{enumerate}
    \item We propose the FMG-Net and W-Net architectures for 3D medical imaging segmentation. These architectures are based on the well know FMG and W-cycle multigrid methods for solving linear systems of equations. 

    \item We show that our proposed architectures achieve superior segmentation performance vs. the popular U-Net architecture and that each architecture converges to lower loss values in fewer epochs. 
\end{enumerate}

\section{Materials and Methods}
\label{sec:methods}

\subsection{FMG-Net and W-Net}
\label{sec:methods-mg-nets}
\begin{figure}[ht!]
    \centering
    \includegraphics[width=0.55\textwidth, height=2.00in]{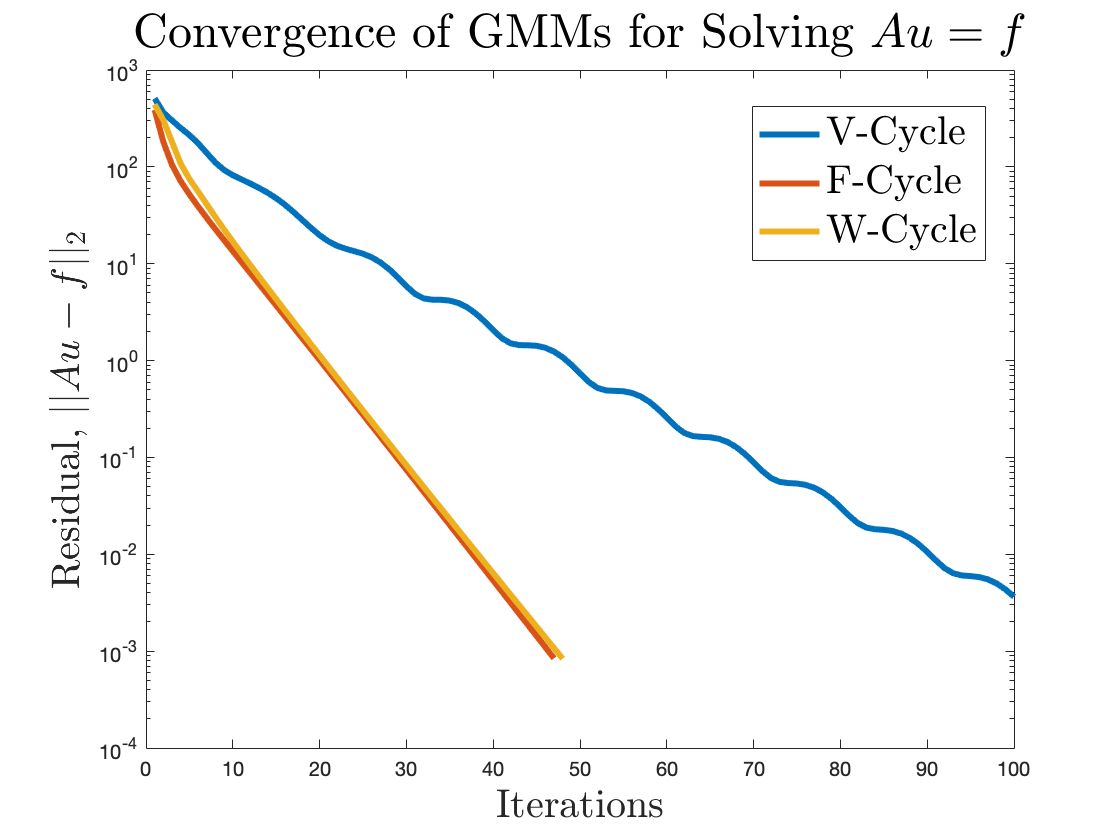}
    \caption{Convergence of V-cycle, FMG-cycle, and W-cycle GMMs. The linear system $Au = f$ results from the finite difference discretization of the Poisson problem. In this figure, we see that the FMG and W-cycles converge to lower residual values in fewer iterations than the V-cycle. \label{fig:gmm-convergence}}
\end{figure}

The V-cycle, W-cycle, and FMG-cycle are different implementations of GMMs, with varying complexities and convergence properties. The V-cycle is the simplest and a commonly used multigrid method, where a sequence of smoothing and restriction operations are performed on the system at multiple grid levels, followed by a sequence of prolongation operations with corrections (i.e., skip connections) to interpolate the solution back to the finest grid. The W-cycle and FMG-cycle are more complex and computationally intensive, but they offer superior convergence properties by performing additional smoothing, interpolation, and correction operations at intermediate grid levels \cite{multigrid, saad2003iterative}. Figure \ref{fig:gmm-convergence} illustrates the faster convergence of the FMG and W-cycles vs. the V-cycle for solving a linear system of equations resulting from the finite difference discretization of the Poisson problem $-\Delta u = f$ with zero Dirichlet boundary conditions. 

\begin{figure}[ht!]
    \centering
    \includegraphics[width=0.9\textwidth]{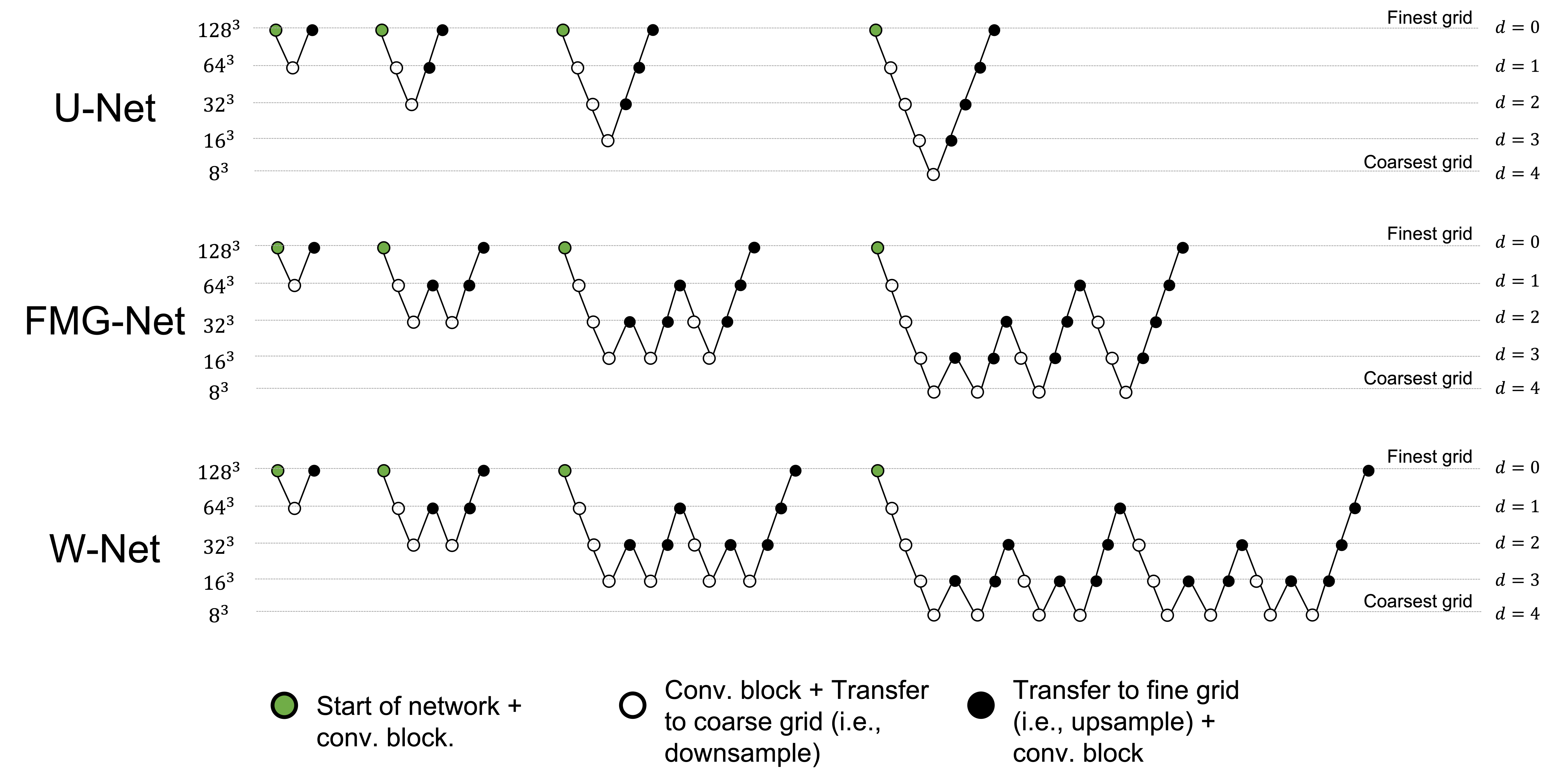}
    \caption{Sketches of U-Net, W-Net, and FMG-Net for different depths. \label{fig:mg-nets-depth}}
\end{figure}

Given that the U-Net architecture and the V-cycle are nearly identical, we propose the FMG-Net and W-Net architectures which mirror the structure of the FMG and W-cycles, respectively. We hypothesize that these architectures, with more complex interactions between features at intermediate grid levels than the U-Net, can achieve superior segmentation performance and faster convergence to lower loss function values. Figure \ref{fig:mg-nets-depth} sketches the U-Net, FMG-Net, and W-Net architectures for various network depths. Note that we define the depth of each network to be the number of different resolution grids used by each architecture. The following rules govern the skip connections in each of these networks:

\begin{enumerate}
    \item If we transfer features to a coarser grid (i.e., are on an encoder branch), then we pass all of the features to all subsequent upsampling operations at the current grid level.

    \item If we encounter a ``peak'' (i.e., a transfer to a finer grid immediately followed by a transfer to a coarser grid), then we pass the features from the peak only to the next set of features at the same grid level.
\end{enumerate}

\begin{figure}[ht!]
    \centering
    \includegraphics[width=0.75\textwidth]{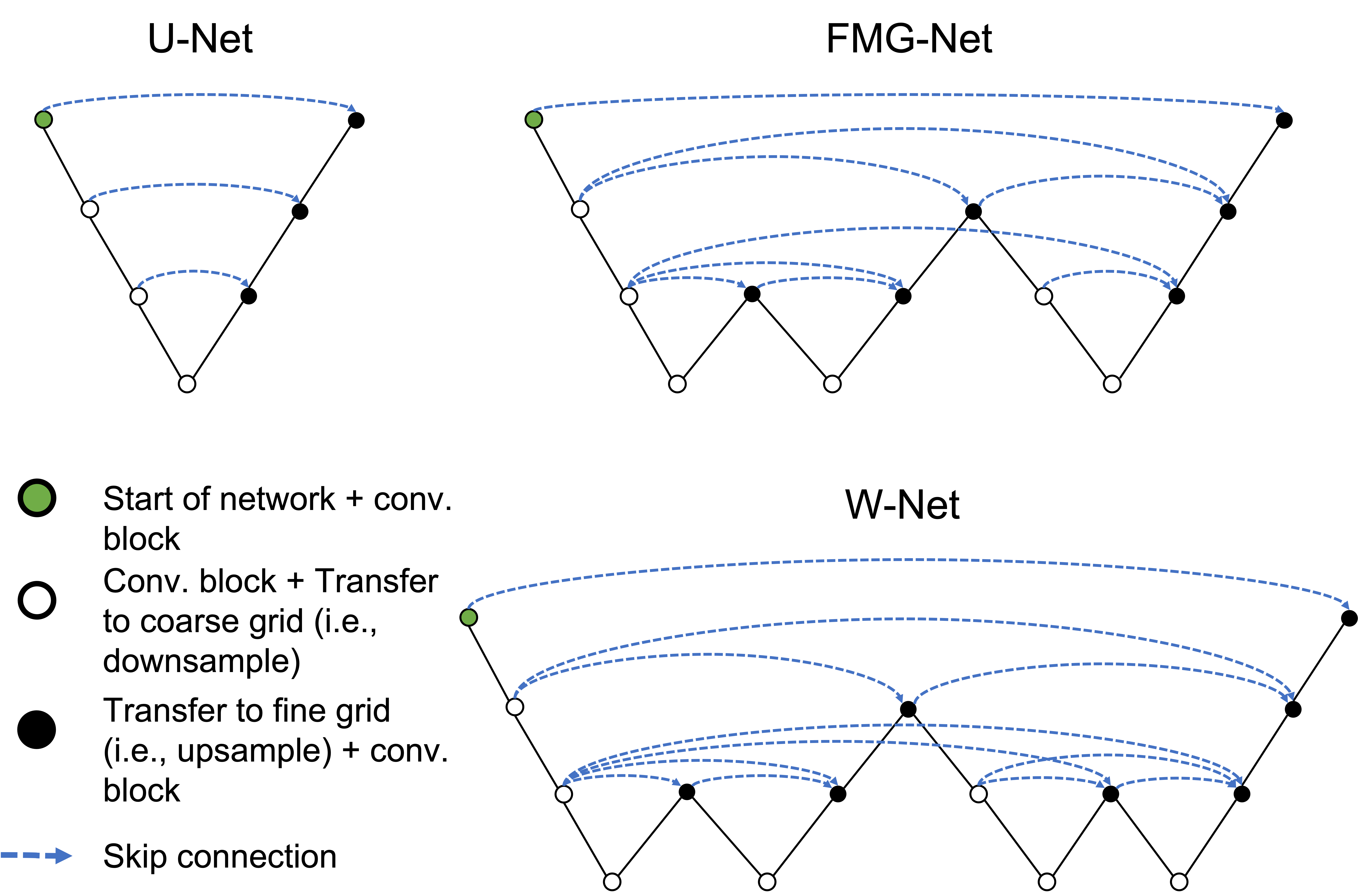}
    \caption{Sketches of U-Net, W-Net, and FMG-Net with skip connections. \label{fig:mg-nets-skip}}
\end{figure}

Figure \ref{fig:mg-nets-skip} illustrates these skip connection patterns for the U-Net, FMG-Net, and W-Net for the network depth set to three. In addition, each convolutional block (i.e. conv. block) consists of two convolutions, each followed by batch normalization and a ReLU non-linearity. Finally, we use the PocketNet paradigm proposed by Celaya et al. for the FMG-Net and W-Net, and \emph{do not} double the number of features after downsampling to a coarser grid.

\subsection{Data}
\label{sec:data}
We test our proposed network architectures on multi-label tumor segmentation in the publicly available (via CC BY 3.0 license) MICCAI Brain Tumor Segmentation (BraTS) Challenge 2020 dataset \cite{brats1, brats2, brats3}. The BraTS training set contains 369 anonymized multimodal scans from 19 institutions. Each set of scans includes a T1-weighted, post-contrast T1-weighted, T2-weighted, and T2 Fluid Attenuated Inversion Recovery volume along with a multi-label ground truth segmentation. The annotations include the GD-enhancing tumor (ET - label 4), the peritumoral edema (ED - label 2), and the necrotic and non-enhancing tumor core (NCR/NET - label 1). The final segmentation classes are the whole tumor (WT - labels 1, 2, and 4), tumor core (TC - labels 2 and 4), and ET. All volumes are provided at an isotropic voxel resolution of 1$\times$1$\times$1 mm$^3$, co-registered to one another, and skull stripped, with a size of 240$\times$240$\times$155. We crop each image according to the brainmask (i.e., non-zero voxels) and apply z-score intensity normalization on only non-zero voxels for pre-processing. The BraTS training dataset is available for download at \url{https://www.med.upenn.edu/cbica/brats2020/registration.html}. 

\subsection{Metrics}
\label{sec:metrics}
We used three common medical imaging segmentation metrics to evaluate the performance of the proposed FMG-Net and W-Net architectures - the Dice coefficient, 95th percentile Hausdorff distance, and average surface distance. These metrics are implemented using the SimpleITK Python package \cite{simpleitk1, simpleitk2, simpleitk3}.

\emph{Dice Coefficient} - The Dice coefficient is a widely used metric for evaluating the similarity between the predicted segmentation and ground truth \cite{dice}. It measures the overlap between the two sets and ranges from 0 to 1, where a value of 1 indicates a perfect match between the predicted and ground truth segmentations.

\emph{95th Percentile Hausdorff Distance} - While the Dice score is a commonly used metric for comparing two segmentation masks, it is not sensitive to local differences, as it represents a global measure of overlap. Therefore, we compute a complimentary metric, the 95th percentile Hausdorff distance (HD95), which is a distance metric that measures the maximum of the minimum distances between the predicted segmentation and the ground truth at the 95th percentile. The HD95 is a non-negative real number measured in millimeters, with a value of 0mm indicating a perfect prediction.

\emph{Average Surface Distance} - The average surface distance (ASD) measures the average distance between the predicted segmentation and the ground truth along the surface of the object being segmented. It is a more fine-grained metric than the HD95 since it captures the surface-level details of the segmentation and can provide insight into the quality of the segmentation. The ASD is a non-negative real number measured in millimeters with a perfect prediction achieving a value of 0mm.

\subsection{Training and Testing Protocols}
\label{sec:training}
We compare the segmentation performance for the popular U-Net architecture vs. the proposed FMG-Net and W-Net architectures described in Section \ref{sec:methods-mg-nets}. We select the Dice loss with cross-entropy for each architecture as a loss function \cite{nnunet}. We initialize the first layer in each network with 32 feature maps and use the Adam optimizer \cite{adam}. The learning rate is set to 0.0003. A five-fold cross-validation scheme is used to train and evaluate each loss function. That is, We train on 80\% of the images for each fold and generate test-time predictions for the excluded 20\%. We iterate this process until we have test-time predictions for all of the images in our dataset. During training, we select a batch size of two, use a patch size of 128$\times$128$\times$128, and apply the same random augmentation described in \cite{nnunet}. To evaluate the validity of a predicted segmentation mask, we use the metrics described in Section \ref{sec:metrics}. Our models are implemented in Python using TensorFlow (v2.11.0) and trained on an NVIDIA Quadro RTX 8000 GPU \cite{tensorflow, keras}. All network weights are initialized using the default TensorFlow initializers. For reproducibility, we set all random seeds to 42. All other hyperparameters are left at their default values. The code for each network architecture is available at \url{https://github.com/aecelaya/mg-nets}.

\section{Results}
\subsection{Accuracy}
\label{sec:accuracy}
\begin{figure}[ht!]
    \centering
    \includegraphics[width=\textwidth]{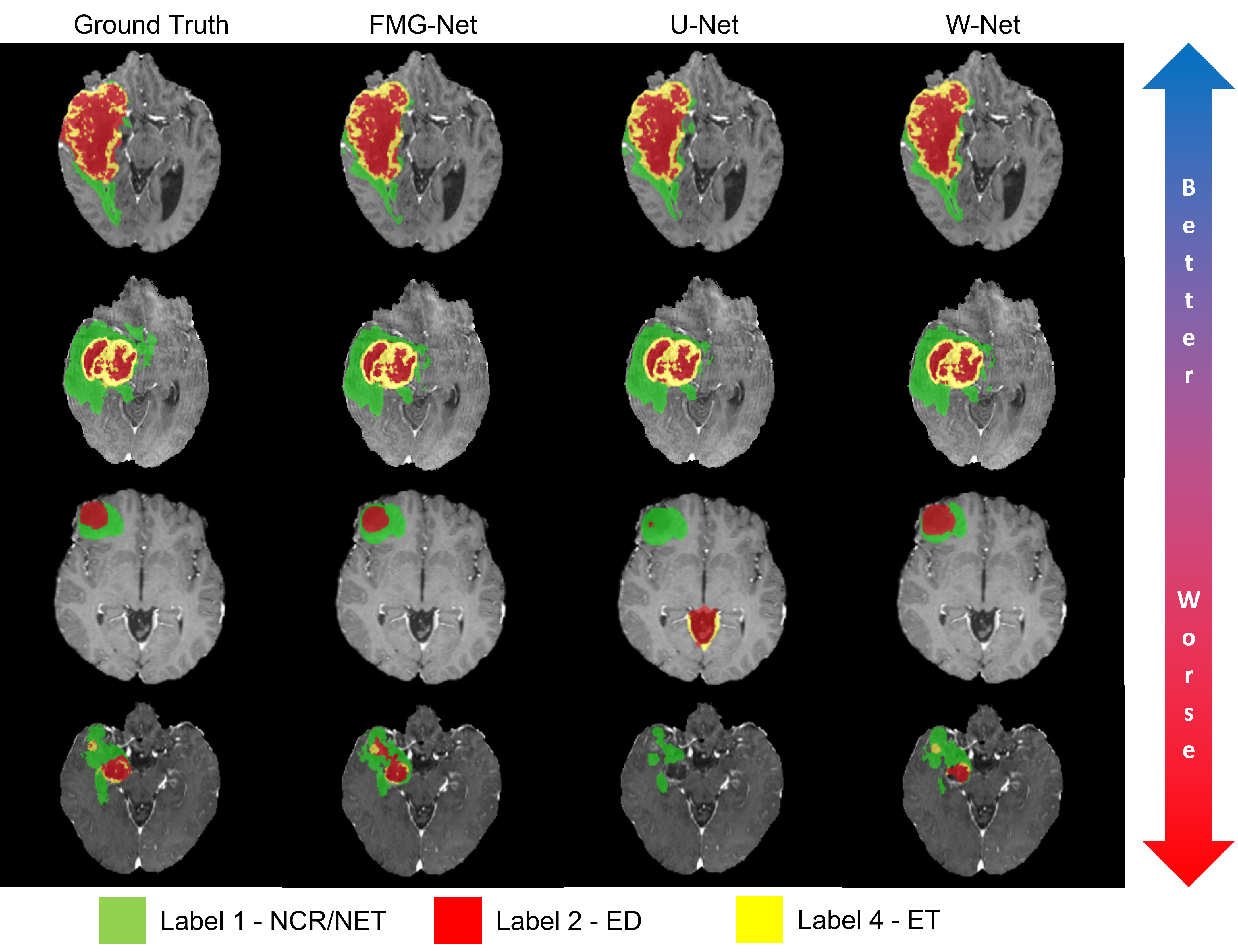}
    \caption{From left to right, ground truth and predictions from the FMG-Net, U-Net, and W-Net for several cases. Here, we see that, even for more difficult cases, the FMG-Net and W-Net can produce visually superior predictions than the U-Net.\label{fig:preds}}
\end{figure}

Using the methods described in Section \ref{sec:methods}, we train a 3D U-Net, FMG-Net, and W-Net and compare their segmentation performance. Table \ref{tab:results} shows the results for each architecture for depths 3, 4, and 5, respectively. In each case, we see that FMG-Net and W-Net generally outperform the U-Net architecture with respect to our selected metrics. Figure \ref{fig:preds} illustrates several predictions from each of the three networks we test. Here, we see that, even for more difficult cases, the FMG-Net and W-Net can produce visually superior predictions than the U-Net.

\begin{table}[ht!]
\centering
\bgroup
\def\arraystretch{1.1}
\resizebox{\textwidth}{!}{%
\begin{tabular}{clccccc}
\hline
Depth & \multicolumn{1}{c}{Architecture} & \# Param. & Class & Dice & Hausdorff 95 (mm) & Avg. Surface (mm) \\ \hline
\multirow{9}{*}{3} & \multirow{3}{*}{U-Net} & \multirow{3}{*}{5,608,036} & WT & 0.8833 (0.1363) & 8.4089 (25.772) & \textbf{2.9406 (19.634)} \\
 &  &  & TC & 0.8240 (0.2156) & 12.265 (41.892) & 6.5104 (38.793) \\
 &  &  & ET & 0.7386 (0.2846) & 37.747 (102.73) & 32.701 (103.15) \\ \cline{2-7} 
 & \multirow{3}{*}{FMG-Net} & \multirow{3}{*}{1,108,356} & WT & \textbf{0.8977 (0.1195)} & \textbf{6.5106 (34.221)} & 4.1923 (33.483) \\
 &  &  & TC & \textbf{0.8422 (0.1940)} & \textbf{8.5238 (39.318)} & \textbf{5.6793 (38.910)} \\
 &  &  & ET & \textbf{0.7647 (0.2650)} & \textbf{26.615 (87.858)} & \textbf{23.501 (88.249)} \\ \cline{2-7} 
 & \multirow{3}{*}{W-Net} & \multirow{3}{*}{1,366,372} & WT & 0.8969 (0.1236) & 7.6357 (43.081) & 6.0217 (43.033) \\
 &  &  & TC & 0.8354 (0.1954) & 9.0972 (43.199) & 6.4143 (43.014) \\
 &  &  & ET & 0.7591 (0.2678) & 28.205 (91.320) & 25.201 (91.791) \\ \hline
\multirow{9}{*}{4} & \multirow{3}{*}{U-Net} & \multirow{3}{*}{22,589,796} & WT & 0.8980 (0.1261) & 4.6759 (27.798) & 3.0767 (27.346) \\
 &  &  & TC & 0.8398 (0.2009) & 10.046 (47.302) & 7.4168 (47.058) \\
 &  &  & ET & 0.7549 (0.2731) & 32.065 (98.065) & 29.197 (98.572) \\ \cline{2-7} 
 & \multirow{3}{*}{FMG-Net} & \multirow{3}{*}{1,862,980} & WT & \textbf{0.9025 (0.1138)} & \textbf{4.3865 (21.538)} & \textbf{2.4595 (20.088)} \\
 &  &  & TC & 0.8378 (0.2023) & 8.3782 (39.088) & 5.6000 (38.693) \\
 &  &  & ET & 0.7610 (0.2695) & \textbf{27.331 (89.607)} & \textbf{24.363 (90.033)} \\ \cline{2-7} 
 & \multirow{3}{*}{W-Net} & \multirow{3}{*}{3,235,684} & WT & 0.9001 (0.1094) & 6.1075 (33.910) & 4.0880 (33.441) \\
 &  &  & TC & \textbf{0.8472 (0.1840)} & \textbf{7.0097 (33.791)} & \textbf{4.3385 (33.429)} \\
 &  &  & ET & \textbf{0.7661 (0.2632)} & 27.764 (91.349) & 25.088 (91.814) \\ \hline
\multirow{9}{*}{5} & \multirow{3}{*}{U-Net} & \multirow{3}{*}{90,500,964} & WT & 0.8968 (0.1129) & 4.7741 (21.625) & 2.2959 (19.467) \\
 &  &  & TC & \textbf{0.8363 (0.1933)} & 6.6805 (28.699) & 3.5758 (27.392) \\
 &  &  & ET & 0.7510 (0.2729) & 30.795 (94.914) & 27.361 (95.237) \\ \cline{2-7} 
 & \multirow{3}{*}{FMG-Net} & \multirow{3}{*}{2,847,652} & WT & 0.8988 (0.0984) & 5.2557 (22.106) & 2.3330 (19.505) \\
 &  &  & TC & 0.8356 (0.1948) & \textbf{6.0468 (21.959)} & \textbf{2.9306 (20.289)} \\
 &  &  & ET & \textbf{0.7525 (0.2707)} & 28.966 (91.423) & 25.663 (91.856) \\ \cline{2-7} 
 & \multirow{3}{*}{W-Net} & \multirow{3}{*}{7,886,692} & WT & \textbf{0.8993 (0.0963)} & \textbf{4.3079 (2.5746)} & \textbf{2.1407 (19.376)} \\
 &  &  & TC & 0.8262 (0.1995) & 7.9477 (34.216) & 4.6273 (33.430) \\
 &  &  & ET & 0.7499 (0.2662) & \textbf{28.893 (91.221)} & \textbf{25.371 (91.757)} \\ \hline
\end{tabular}%
}
\egroup
\vspace{1mm}
\caption{Mean and standard deviation for each metric for the U-Net, FMG-Net, and W-Net architectures with varying depths. For each depth, we highlight the best metric values in bold text. In each case, we see that FMG and W-Nets generally outperform the U-Net architecture regarding our selected metrics. \label{tab:results}}
\end{table}

\subsection{Loss Curve Convergence}
\begin{figure}[ht!]
    \centering
    \includegraphics[width=\textwidth]{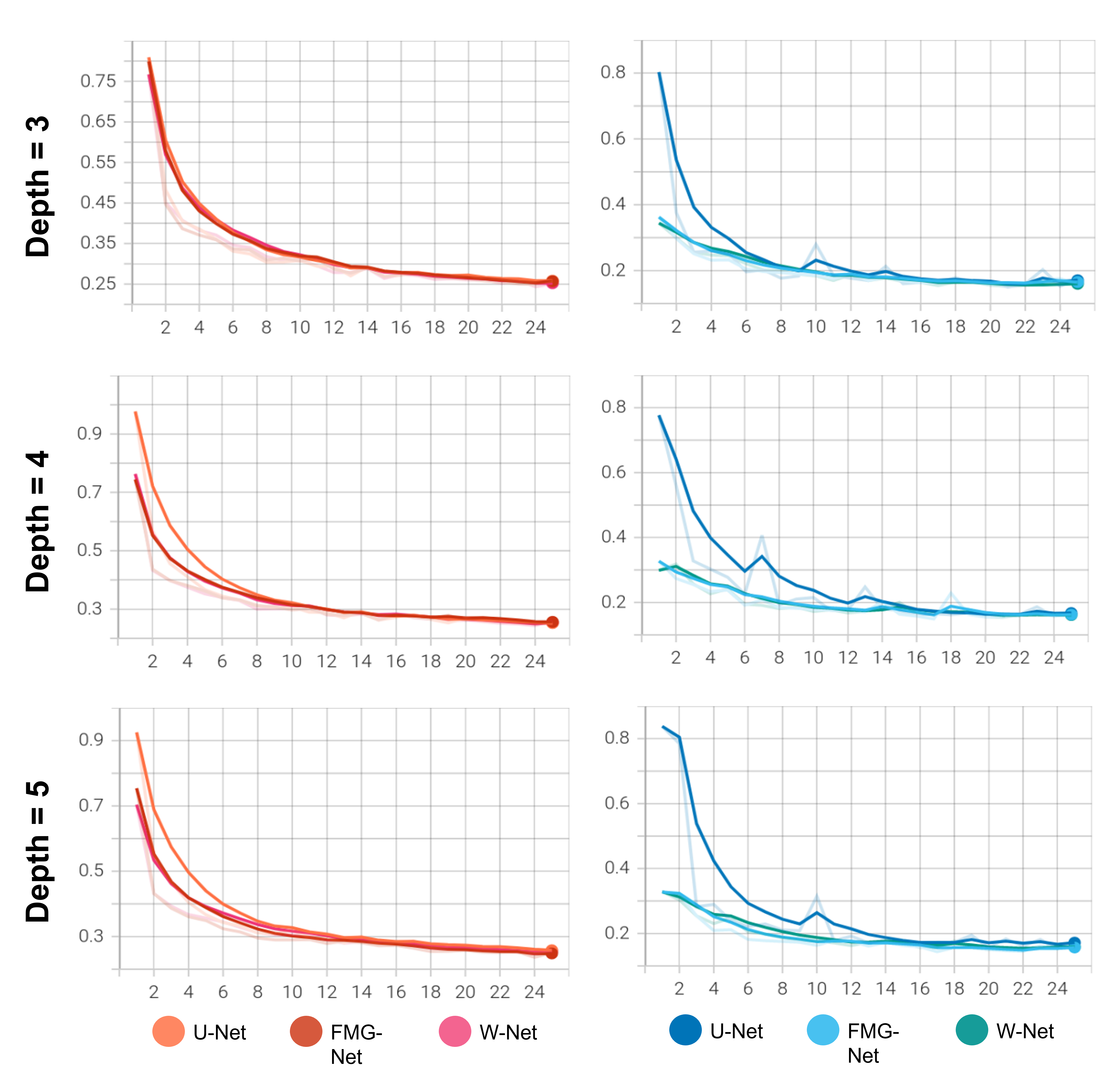}
    \caption{(Left) Training loss curves for depths 3 through 5 for the U-Net, FMG-Net, and W-Net architectures for the first 25 training epochs. (Right) Validation loss curves for depths 3 through 5 for the U-Net, FMG-Net, and W-Net architectures for the first 25 training epochs. As the depth increases, the separation between the loss curves for the U-Net and our proposed architectures becomes more apparent. Especially in the validation curves, our architectures achieve lower loss values in fewer epochs than the U-Net. \label{fig:loss-curves}}
\end{figure}

We also compare the loss curves (training and validation) for varying depths for the U-Net, FMG-Net, and W-Net architectures. Figure \ref{fig:loss-curves} shows the loss curves for each network for depths 3 through 5. Although each network converges to similar loss values, the FMG-Net and W-Net architectures generally do so in fewer epochs. This faster convergence (in terms of epochs) is especially true for the validation loss curves.

\section{Discussion}
\label{sec:discussion}
The results presented above show that our proposed FMG and W-Net, which incorporate the design of two well-known corresponding GMM schemes, can achieve superior segmentation accuracy compared to the widely used U-Net architecture on the challenging BraTS dataset. Indeed, the proposed architectures seem to help address current CNNs' limitations in handling multi-scale features. This ability to improve upon handling large and fine-scale features is evident in the FMG and W-Net's superior segmentation accuracy for the various tumor subcomponents (which vary significantly in size and shape) in the BraTS dataset compared to the U-Net architecture. Figure \ref{fig:preds} makes this clear visually, with the FMG and W-Net architectures resolving finer scale features in smaller, more difficult-to-segment tumors. However, one of the objectives of future work will be to add more comparisons to other architectures like nnUNet and HRNet (non-U-shaped) \cite{hrnet}. Additionally, incorporating an adaptive framework like the nnUNet (i.e., depth and network shape based on the dataset) into our architectures may improve our segmentation accuracy.

It is clear from Table \ref{tab:results} that for depth equal to three, the FMG-Net outperforms the U-Net and W-Net. It is also important to note that even in that case, the W-Net yields more accurate results than the U-Net. However, the picture is less clear regarding the FMG vs. the W-Net architecture for depths four and five. In these cases, the FMG and W-Net still generally outperform the U-Net, but it is unclear whether one is more accurate. Additionally, there is no trend in the accuracy for these networks vs. the depth. This behavior is not only consistent with multigrid theory, which does not guarantee faster convergence to a solution as a function of depth \cite{multigrid, shah1989analysis}, but also with the results in \cite{jena2023analysis}, which show that for the BraTS dataset, the depth of the U-Net architecture does not significantly affect accuracy. However, this may not be true for other datasets like the Liver and Tumor Segmentation or Medical Segmentation Decathalon datasets \cite{lits, antonelli2022medical}. Further testing on these is needed to validate our results further, determine which cases to use the FMG or W-Net architectures, and asses the effect of the depth parameter on accuracy.

Because the FMG and W-Net architectures use substantially fewer parameters than the U-Net, they offer advantages regarding speed-ups in the average time per training step and memory usage. The FMG-Net offers the highest training speed-ups vs. the U-Net, with reductions in the average time per training step ranging from 7\% to 20\%, depending on depth. The W-Net offers more modest speed-ups, with reductions in the average time per training step ranging from 2\% to 6\%, depending on depth. The W-Net architecture not only has more parameters than the FMG-Net but also requires more upsampling layers, downsampling layers, and skip connections (i.e., concatenations) than the FMG-Net, which adds to the computational cost of the architecture. We see similar results from the point of view of GPU memory usage, with the FMG-Net offering the most savings (roughly 20\% vs. the U-Net) and the W-Net giving more modest savings (roughly 8\% vs. the U-Net). While further work is needed to determine which architecture results in higher accuracy, it is clear that the FMG-Net produces the highest savings in terms of time and memory requirements. Practitioners considering the use of our proposed architectures should also consider this advantage in computational cost. 

In addition to computational cost savings offered by our architectures, we also see that the loss curves for the FMG and W-Net achieve lower loss values in fewer epochs than the U-Net. Indeed, as the depth increases, the separation between the loss curves for the U-Net and our proposed architectures becomes more apparent. Especially in the validation curves, our architectures achieve lower loss values in fewer epochs than the U-Net. This convergence behavior is consistent with the convergence for the V-cycle vs. the FMG and W-cycles seen in Figure \ref{fig:gmm-convergence}. 

The FMG-Net and W-Net achieve higher accuracy than the U-Net architecture while using substantially fewer parameters. One possible explanation is that our proposed architectures rely on more complex interactions between features at multiple resolutions. These results suggest that the success of CNNs for medical imaging segmentation may have less to do with the number of parameters in each architecture but rather the richness of the interactions of features at different grid levels. Further testing on different datasets is necessary to validate our findings' generalizability and explore the potential of FMG-Net and W-Net in other medical imaging applications. Overall, our study highlights the promising potential of incorporating GMM principles into CNNs to improve the accuracy and efficiency of medical imaging segmentation.

\begin{ack}
The Department of Defense supports Adrian Celaya through the National Defense Science \& Engineering Graduate Fellowship Program. David Fuentes is partially supported by R21CA249373. Beatrice Riviere is partially supported by NSF-DMS2111459. This research was partially supported by the Tumor Measurement Initiative through the MD Anderson Strategic Research Initiative Development (STRIDE), NSF-2111147, and NSF-2111459.
\end{ack}




\bibliographystyle{ieeetr}
{\small
\bibliography{sources}}

\end{document}